\documentclass[twocolumn]{biophys-new}
\usepackage[utf8]{inputenc}
\usepackage{graphicx}
\usepackage[colorlinks,allcolors=cyan!70!black]{hyperref}
\usepackage{color}

\usepackage{lipsum}
\usepackage{amssymb}
\usepackage{gensymb}

\newcommand{\beginsupplement}{%
        \setcounter{table}{0}
        \renewcommand{\thetable}{S\arabic{table}}%
        \setcounter{figure}{0}
        \renewcommand{\thefigure}{S\arabic{figure}}%
     }

\title{Membrane Morphologies Induced by Arc-Shaped {Scaffolds} are Determined by Arc Angle and Coverage}
\runningtitle{} 
\date{}

\author{Francesco Bonazzi and Thomas R.\ Weikl \hspace*{3cm} \\ \small Max Planck Institute of Colloids and Interfaces, Department of Theory and Bio-Systems, Science Park Golm, 14424 Potsdam, Germany \hspace*{4cm} }
\runningauthor{} 

\begin{document}

\begin{frontmatter}

\begin{abstract}
 The intricate shapes of biological membranes such as tubules and membrane stacks are induced by proteins. In this article, we systematically investigate the membrane shapes induced by arc-shaped { scaffolds} such as proteins and protein complexes with coarse-grained modeling and simulations. We find that arc-shaped { scaffolds} induce membrane tubules at membrane coverages larger than a threshold of about 40\%, irrespective of their arc angle. The membrane morphologies at intermediate coverages below this tubulation threshold, in contrast, strongly depend on the arc angle. { Scaffolds} with arc angles of about $60\degree$ {akin to} N-BAR domains do not change the membrane shape at coverages below the tubulation threshold, while { scaffolds} with arc angles larger than about $120\degree$ induce double-membrane stacks at intermediate coverages. The { scaffolds} stabilize the curved membrane edges that connect the membrane stacks, as suggested for complexes of reticulon proteins.  Our results provide general insights on the determinants of membrane shaping by arc-shaped { scaffolds} . 
\end{abstract}

\end{frontmatter}

\section*{Introduction}

The shapes of biological membranes that surround cells and cellular organelles are often highly curved \cite{Shibata09,Kozlov14,Nixon16}. The membrane curvature is induced and regulated by proteins such as the arc-shaped BAR domains \cite{Takei99,Peter04,Rao11,Baumgart11,Mim12b} or the reticulons \cite{Voeltz06,Zurek11}, which have been suggested to oligomerize into arc-shaped protein { complexes} \cite{Shibata08}. Arc-shaped proteins and protein { complexes} can induce membrane tubules \cite{Frost08,Hu08,Daum16}, but have also been associated with other highly curved membrane structures. Reticulon proteins, for example, are involved in the generation of the membrane tubules of the endoplasmic reticulum (ER) and have been suggested to stabilize the highly curved edges \cite{Shibata10,Schweitzer15b} that connect stacked membrane sheets of the ER \cite{Terasaki13,Nixon16}. Electron microscopy indicates that BAR domain proteins can form highly ordered helical coats around membrane tubules  \cite{Frost08,Mim12a,Adam15} that are apparently held together by specific protein-protein interactions, as well as rather loose, irregular arrangements  
\cite{Daum16}. 
{  
The variability of distances and angles between neighboring BAR domains in these loose arrangements suggests that the arrangements form without specific protein-protein interactions \cite{Daum16} and, thus, may be dominated by membrane-mediated interactions between the proteins \cite{Weikl18,Phillips09,Reynwar07}. 
These indirect, membrane-mediated interactions arise because the overall bending energy of the membrane depends on the distance and orientation of curvature-inducing proteins.} In simulations with coarse-grained models, a variety of morphologies with tubular or disk-like membrane shapes have been observed \cite{Ayton07,Ayton09,Ramakrishnan13,Tourdot14,Noguchi15,Noguchi16,Noguchi16b}. The disk-like shapes consist of a double-membrane stack connected by a curved edge and are counterparts of the connected, stacked membrane sheets in the much larger membrane systems investigated in experiments \cite{Terasaki13,Nixon16}. 

In this article, we systematically investigate the membrane morphologies induced by arc-shaped { scaffold} particles such as proteins or protein complexes with coarse-grained modeling and simulations. In our coarse-grained model of membrane shaping, the membrane is described as a triangulated elastic surface, and the particles as segmented arcs that induce membrane curvature by binding to the membrane. { The direct particle-particle interactions are purely repulsive and only prevent particle overlap. The particle arrangements in our model are therefore governed by indirect, membrane-mediated interactions. 
These particle arrangements are essentially unaffected by the membrane discretization because the particles are not embedded in the membrane, in contrast to previous elastic-membrane models. In previous models, curvature-inducing particles have been described as nematic objects embedded on the vertices of a triangulated membrane \cite{Ramakrishnan13,Tourdot14}, or as curved chains of beads embedded in a two-dimensional sheet of beads that represents the membrane \cite{Noguchi15,Noguchi16}.}

Our main aim here is to obtain general classification of the membrane morphologies induced by arc-shaped scaffold particles that do not exhibit specific attractive interactions. This classification is obtained from simulations in which the overall number of particles exceeds the number of membrane-bound particles. The membrane coverage then is not constrained by the number of available particles, which leads to rather sharp transitions between `pure' spherical, tubular, or disk-like morphologies in our simulations. Previous elastic membrane models, in contrast, have been investigated for a fixed number of membrane-embedded or bound particles, which typically leads to `mixed' membrane morphologies, e.g.~morphologies with membrane tubules or disks protruding from a spherical versicle.

We find that the membrane shape is fully determined by the arc angle and the membrane coverage of the particles. For all considered arc angles of the particles between $60\degree$ and $180\degree$, membrane tubules are formed at particle coverages that exceed about 40\%. Arc angles of $60\degree$ roughly correspond to the angle enclosed by BAR domain proteins such as the Arfaptin BAR domain and the endophilin and amphiphysin N-BAR domains \cite{Qualmann11,Masuda10}, while larger arc angles up to $180\degree$ have been postulated for reticulon scaffolds \cite{Shibata10,Schweitzer15b}. At smaller membrane coverages below 40\%, particles with arc angles of about  $60\degree$ do not change the membrane morphologies in our model, while particles with arc angles larger than about 120$\degree$ induce disk-like double-membrane stacks by stabilizing curved edges. Particles with arc angles around $90\degree$ lead to faceted, irregular membrane morphologies at smaller coverages. { The arrangements of particles with arc angles of $60\degree$ along tubules in our simulations is similar to the rather loose arrangement of N-BAR domains observed in electron microscopy experiments \cite{Daum16}. This similarity supports the suggestion that these rather loose arrangements of N-BAR domains are dominated by membrane-mediated interactions.}

\section*{Methods}

%
\begin{figure}[t!]
\centering
\includegraphics[width=\linewidth]{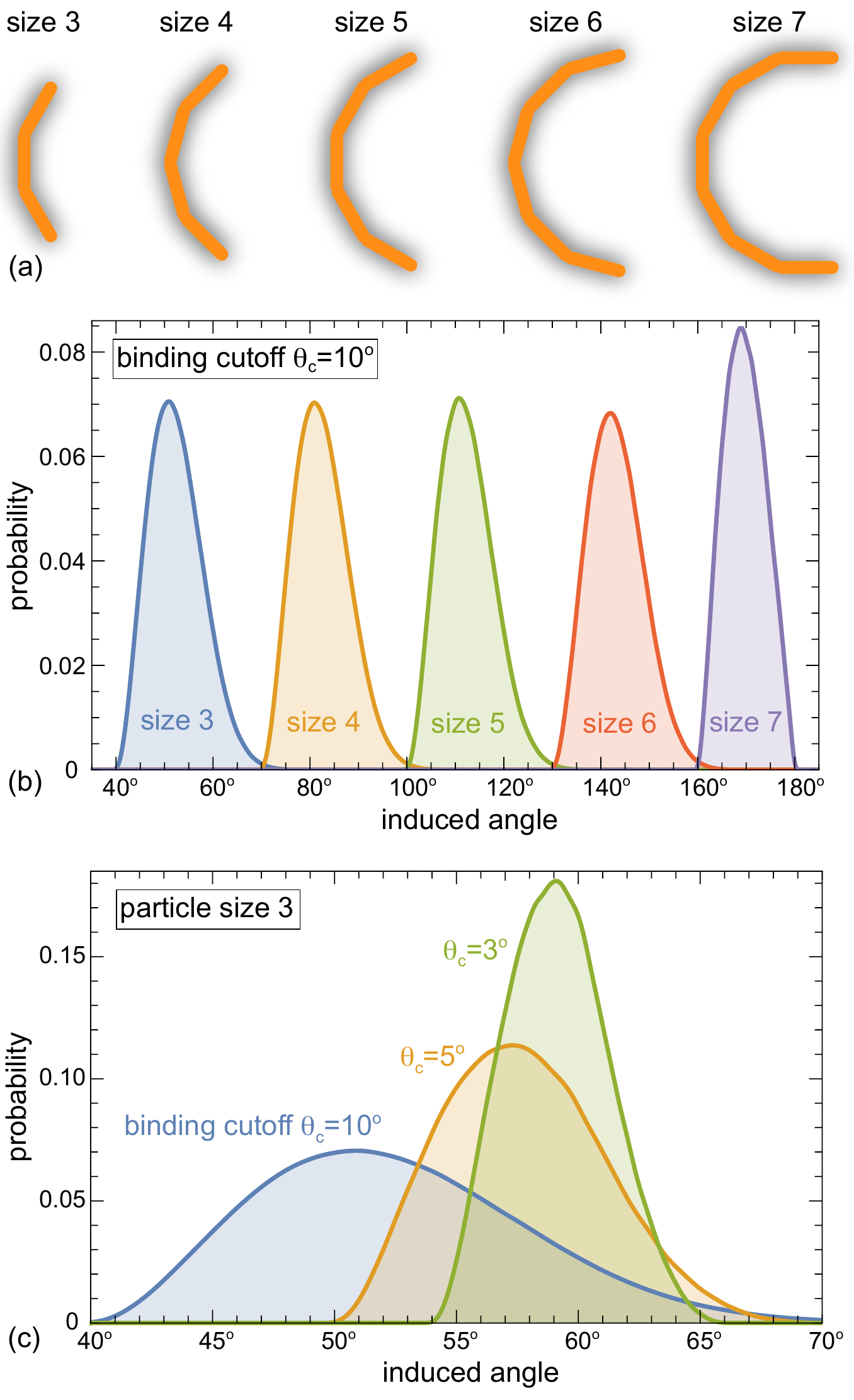}
\caption{(a) Arc-shaped particles composed of 3 to 7 planar segments with an angle of $30\degree$ between neighboring segments. The angle between the two end segments of the particles is $60\degree$ for particles of size 3, and  $90\degree$,  $120\degree$, $150\degree$, and $180\degree$ for particles of size 4, 5, 6, and 7, respectively. (b) Distributions of angles between the membrane triangles that are bound to the end segments of the particles for the binding cutoff $\theta_c = 10\degree$ of the particle-membrane adhesion potential (see Equation (\ref{Vpm})). The mean values of these angle distributions are $52.5\degree$, $82.6\degree$, $112.4\degree$, $143.2\degree$, and $169.7\degree$ for particles of size 3 to 7, respectively. (c) Distributions of angles between the membrane triangles bound to the end segments of particles of size 3 for different values of the binding cutoff $\theta_c$. The mean value of the distributions increases from $52.5\degree$ for $\theta_c=10\degree$ to 57.8\degree for $\theta_c=5\degree$ and $59.1\degree$ for $\theta_c=3\degree$. 
}
\label{figure-1}
\end{figure}

\subsection*{Model}
We model the membrane as a discretized closed surface. The bending energy of a closed continuous membrane without spontaneous curvature is the integral ${\cal E}_\text{be} = 2\kappa \oint M^2 \,dS$ over the membrane surface with local mean curvature $M$ \cite{Helfrich73}. We use the standard discretization of the bending energy for triangulated membrane described in Ref.\ \cite{Julicher96,Bahrami12} and choose as typical bending rigidity the value $\kappa = 10 k_B T$ \cite{Dimova14}. Our discretized membranes are composed of either $n_t = 2000$ triangles or $n_t = 5120$ triangles. The edge lengths of the triangles are kept within an interval $[a_m, \sqrt{3} a_m]$ \cite{Julicher96,Bahrami12}. The area of the membrane is constrained to ensure the near incompressibility of lipid membranes \cite{Lipowsky05}. The strength of the harmonic constraining potential is chosen such that the fluctuations of the membrane area are limited to less than $0.5\%$. The enclosed volume is unconstrained to enable the full range of membrane morphologies. Coarse-grained molecular simulations indicate that the full spectrum of bending fluctuations can be described for a membrane discretization length $a_m$ of about 5 nm\cite{Goetz99}. 

Our arc-shaped particles are composed of 3 to 7 identical planar quadratic segments. Neighboring segments share a quadratic edge and enclose an angle of $30\degree$  in most of our simulations (see Figure 1(a)). The arc angle of the particles, i.e.\ the angle between the first and last segment, then adopts the values $60\degree$, $90\degree$, $120\degree$, $150\degree$, and $180\degree$ for particles with 3, 4, 5, 6, and 7 segments, respectively. In addition, we consider particles composed of 5 segments with an angle of $15\degree$ between neighboring segments. These particles have the same arc angle of $60\degree$ as particles composed of 3 segments with angle $30\degree$ between neighboring segments, but have a larger size and smaller curvature compared to the particles of 3 segments.

Each planar segment of a particle interacts with the nearest triangle of the membrane via the particle-membrane adhesion potential
\begin{equation}
V_\text{pm} = - U f_r(r) f_\theta(\theta),
\label{Vpm}
\end{equation}
where $r$ is the distance between the center of the segment and the center of the nearest triangle, $\theta$ is the angle between the normals of the segments and this triangle, and $U$ is the adhesion energy per particle segment. The distance-dependent function $f_r$ is a square-well function that adopts the values $f_r(r) = 1$ for $r_1 < r < r_2$ and $f_r(r)=0$ otherwise. The angle-dependent function $f_\theta$ is square-well function with values $f_\theta(\theta) = 1$ for $|\theta| < \theta_c$ and $f_\theta(\theta) = 0$ otherwise. By convention, the normals of the membrane triangles are oriented outward from the enclosed volume of the membrane, and the normals of the particle segments are oriented away from the center of the particle arc. The particles then bind to the membrane with their inner, concave side that is oriented towards the center of the arc. We use the parameter values $r_1 = 0.25 a_m$ and $r_2 = 0.75 a_m$ in all our simulations, and the value $\theta_c = 10\degree$ in most of our simulations. In simulations with the smallest particles composed of 3 segments, we consider also the values $\theta_c = 3\degree$ and $5\degree$, besides $\theta_c = 10\degree$. The overlapping of particles is prevented by a purely repulsive hard-core interaction that only allows distances between the centers of the planar segments of different particles that are larger than $a_p$. The hard-core area of a particle segment thus is $\pi a_p^2/4$. We use this hard-core area in calculating the membrane coverage of bound particles. We choose the value $a_p = 1.5 a_m$ for the linear size of the planar particle segments. The particle segments then are slightly larger than the membrane triangles with minimum side length $a_m$, which ensures that different particle segments bind to different triangles.\\

\begin{figure*}[hbt!]
\centering
\includegraphics[width=0.95\linewidth]{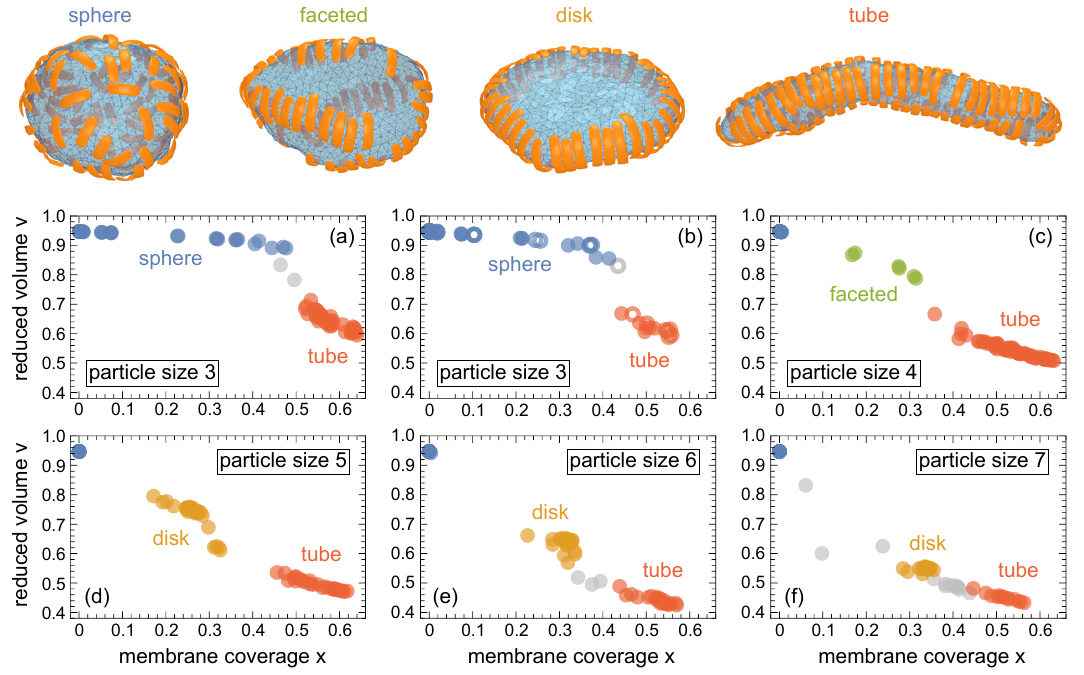}
\caption{(a) Reduced volume versus membrane coverage for membrane morphologies induced by (a) particles of size 3 with binding cutoff $\theta_c = 10\degree$, (b) particles of size 3 with binding cutoffs $\theta_c = 3\degree$ (full circles) and $\theta_c = 5\degree$ (open circles), and (c) to (f) particles of size 4 to 7 with binding cutoff $\theta_c = 10\degree$.  Clusters of points with the same morphology are classified as spheres (blue), faceted (green), disks (orange), and tubes (red). The grey points in (b) and (c) are intermediate between spherical and tubular, the grey points in (e) are intermediate between disk-like and tubular, and the grey points in (f) correspond to strongly metastable morphologies in which only a part of the membranes is tubular and covered with particles. The data result from simulations with membranes composed of 2000 triangles. In these simulations, the membrane area is constrained to the average value $A \simeq 0.677 n_t a_m^2 \simeq 1.35 \cdot 10^3 a_m^2$ where $a_m$ is the minimum edge length of the triangulated membrane. The overall number of particles in our  simulations is either $N=200$ or $400$, and the adhesion energy per particle segment adopts one of the values $U = 3, 4, 5, \ldots 20\, k_B T$ (see Figure S2). For each combination of particle number $N$ and adhesion energy $U$, we have performed simulations starting from an initial spherical, disk-like, or tubular membrane shape (see Figure S1). The membrane and particles are enclosed in a cubic box of volume $V_\text{box} \simeq 1.26\cdot 10^5 a_m^3$. This box volume is 27 times larger than the volume of a perfect sphere with the membrane area $A$ given above. 
}
\label{figure-2}
\end{figure*}

\subsection*{Simulations}
We have performed Metropolis Monte Carlo simulations in a cubic box with periodic boundary conditions. The simulations consist of four different types of Monte Carlo steps: membrane vertex translations, membrane edge flips, particle translations, and particle rotations. Vertex translations enable changes of the membrane shape, while edge flips ensure  membrane fluidity \cite{Gompper97}. In a vertex translation, a randomly selected vertex of the triangulated membrane is translated along a random direction in three-dimensional space by a distance that is randomly chosen from an interval between $0$ and $0.1 a_m$. In a particle translation, a randomly selected particle is translated in random direction by a random distance between $0$ and $a_m$. In a particle rotation, a randomly selected particle is rotated around a rotation axis that passes trough the central point along the particle arc. For particles that consist of 3, 5, or 7 segments, the rotation axis runs through the center of the central segments. For particles of 4 or 6 segments, the rotation axis runs through the center of the edge that is shared by the two central segments. The rotation axis is oriented in a random direction. The random rotations are implemented using quaternions \cite{Frenkel02,Vesely82} with rotation angles between 0 and a maximum angle of about $2.3\degree$.  Each of these types of Monte Carlo steps occur with equal probabilities for single membrane vertices, edges, or particles. 

The membrane coverage $x$ of the particles in our simulations depends on the overall number of particles and on the adhesion energy $U$ per particle segment. The overall number of particles in our  simulations is either $N=200$ or $400$, and the adhesion energy per particle segment is varied from $U = 3$ to $20\, k_B T$ to obtain the full range of possible coverages. For each combination of particle number $N$ and adhesion energy $U$, we have performed simulations starting from an initial spherical, disk-like, or tubular membrane shape (see Figure S1). The particles are initially randomly distributed in the simulation box outside of the membrane. Simulations starting from initial disk-like and tubular shapes first include only particle translations and rotations to stabilize these initial shapes by bound particles. All simulations then include all four types of MC moves for total simulation lengths between $1\cdot 10^7$ and $8\cdot 10^7$ MC steps per membrane vertex, depending on convergence. To verify convergence, we divide the last $10^7$ MC steps per vertex of a simulation into ten intervals of $10^6$ steps and calculate the average coverage $x$ and reduced volume $v$ of the membrane for each interval. For membranes composed of $n_t = 2000$ triangles,  we take a simulation to be converged if the standard deviations of these ten averages are smaller than 0.02 for both quantities. The values for the  membrane coverage $x$ and reduced volume $v$ given in Figures \ref{figure-2} and \ref{figure-3} are the mean values of these 10 averages. For our larger membranes composed of $n_t = 5120$ triangles, we take a simulation to be converged if the standard deviations of the 10 averages of $x$ and $v$ for the last 10 intervals of $10^6$ MC steps are both smaller than 0.01.

\section*{Results}

The arc-shaped particles of our model induce membrane curvature by binding to the membrane with their inner concave sides. We first consider particles composed of 3 to 7 planar segments with an angle of $30\degree$ between adjacent segments  (see Figure \ref{figure-1}(a)). The arc angle of these particles depends on the particle size, i.e.\ on the number of planar segments. A particle segment is bound to the discretized, triangulated membrane of our model if its distance to the closest membrane triangle is within a given range, and if the particle segment and membrane triangle are nearly parallel with an angle that is smaller than a cutoff angle $\theta_c$ (see Methods for details). The relative area of the particle segments and membrane triangles is chosen such that a particle segment can only be bound to a single membrane triangle. Figures 1(b) and (c) illustrate the distributions of angles between the two membrane triangles that are bound to the two end segments of the particles. These induced membrane angles increase with increasing particle size and with decreasing binding cutoff $\theta_c$. 

The membrane morphologies obtained in our simulations are determined by the size and membrane coverage of the particles (see Figure \ref{figure-2}). The overall number of particles in the simulations is always larger than the number of bound particles covering the membrane. The membrane coverage then depends on the concentration and binding energy of the particles, but is not limited or constrained by the number of available particles. An overlap between particles is prevented by a hard-core repulsion potential. Without particles, the closed membrane of our model adopts a spherical shape because the bending energy of such a membrane vesicle is minimal for the sphere.  For the smallest particles of size 3 and membranes composed of 2000 triangles, the membrane retains a spherical shape until coverages of about 50\% for the binding cutoff angle $\theta_c = 10\degree$ (see Figure \ref{figure-2}(a)), and until coverages of about 45\% for the binding cutoffs $\theta_c =3\degree$ and $5\degree$  (see Figure \ref{figure-2}(b)). At larger coverages, the morphology of the membrane changes from spherical to tubular. This morphology change leads to a drop in the reduced volume $v = 6 \sqrt{\pi} V/A^{3/2}\le 1$, which is a measure for the area-to-volume ratio of the membrane vesicle \cite{Seifert91} and adopts its maximum value of 1 for an ideal sphere. The area $A$ of the membrane is constrained in our simulations to ensure the near incompressibility of lipid membranes \cite{Lipowsky05}, whereas the volume $V$ is unconstrained to allow for the full range of membrane morphologies. 

At intermediate coverages, the membrane morphologies depend on the particle size. For particles of size 4 to 7, spherical morphologies with bound particles do not occur, in contrast to particles of size 3. Instead, particles of size 4 lead to irregular, faceted morphologies at intermediate coverages between about 10\% and 30\%. Particles of size 5 to 7 induce disk-like morphologies at intermediate coverages. However, all particles induce tubular morphologies at sufficiently large coverages.

\begin{figure}[t!]
\centering
\includegraphics[width=\linewidth]{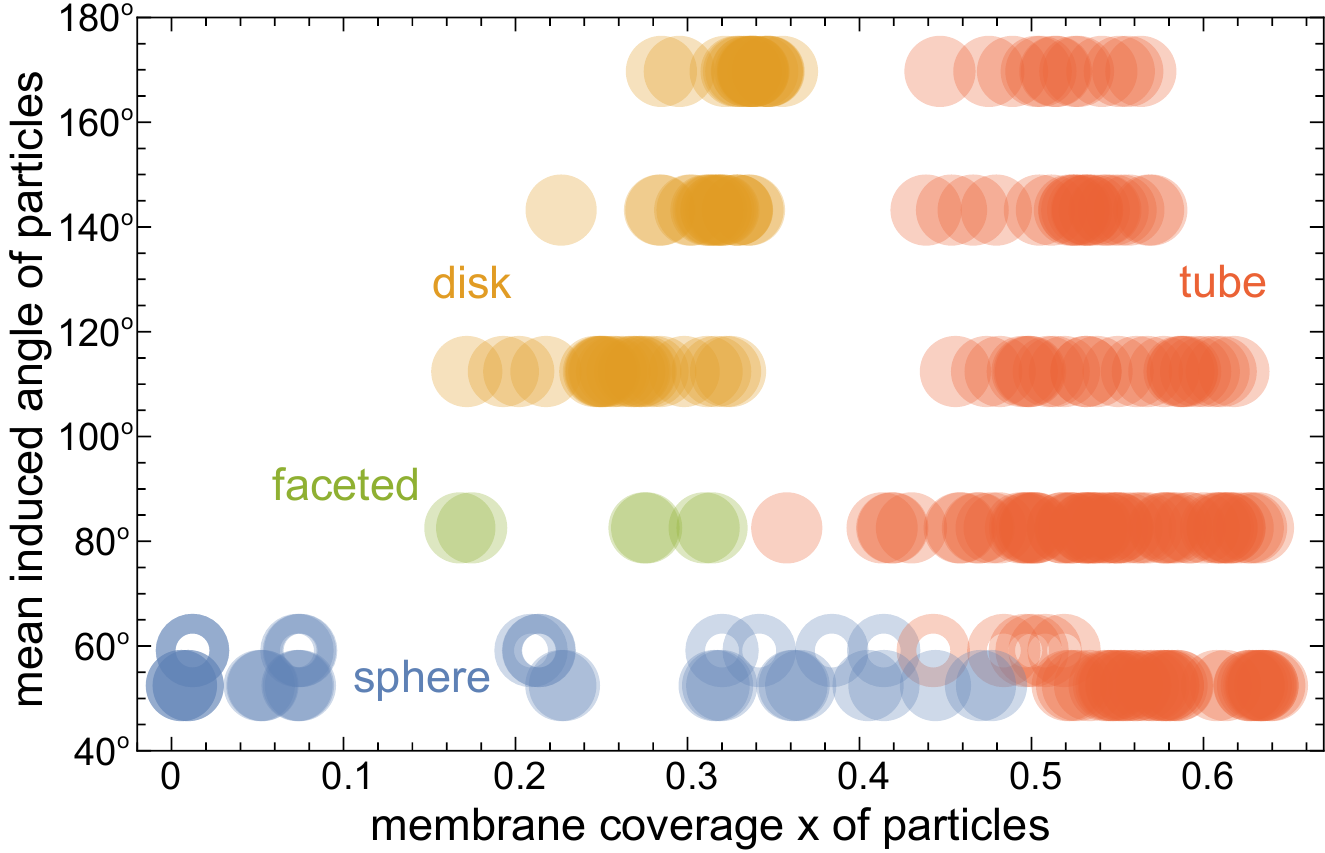}
\caption{Morphology diagram with mean induced angle versus membrane coverage for the data points of Figure \ref{figure-2}. The mean induced angles are the mean values of the angle distributions of Figure \ref{figure-1}. The lines of full circles with the same mean induced angle are from simulations with particles of size 3 to 7 (bottom to top) and binding cutoff $\theta_c = 10\degree$.  The open circles represent the simulation results for particles with size 3 and binding cutoff $\theta_c = 3\degree$. 
}
\label{figure-3}
\end{figure}
\begin{figure*}[bth!]
\centering
\includegraphics[width=0.95\linewidth]{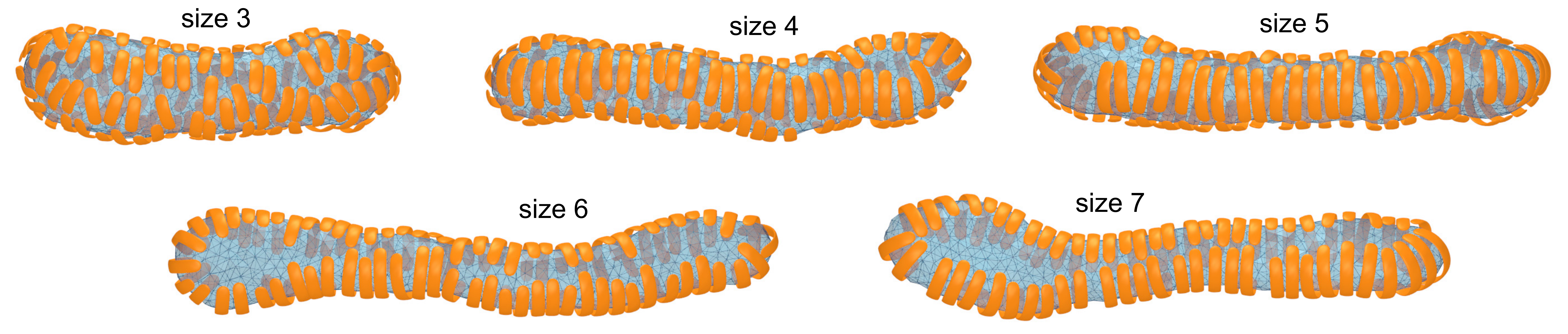}
\caption{Exemplary tubular morphologies for particles composed of 3 to 7 segments with an angle of $30\degree$ between adjacent segments. The membrane coverage is $x = 0.56$ for all morphologies. The membranes consist of 2000 triangles.}
\label{figure-4}
\end{figure*}

The points in Figure \ref{figure-2} result from simulations with different overall particle number, binding energy, or initial membrane shape. The initial membrane shape in our simulations is either spherical, tubular, or disk-like (see Figure S1). Figure \ref{figure-2} only includes points from simulations with converged membrane coverage and reduced volume (see Methods). Simulations that differ in the initial membrane shape may converge to different values for the membrane coverage and reduced volume, which indicates metastabilities. These metastabilities tend to increase with the binding energy and size of the particles. For particles with size 3 to 5, all membrane morphologies emerge in simulations starting from an initially spherical membrane. For particles with size 6 and 7, in contrast, disk-like and tubular morphologies no longer emerge in simulations with an initially spherical membrane because these particles do not bind to spherical membranes within the accessible simulation times. However, all points in the diagrams of figure \ref{figure-2} fall onto a single curve for a given particle size, irrespective of whether these points result from metastable or stable simulations and parameters. These curves imply that the reduced volume $v$ of the membrane is a function of the membrane coverage. For each particle type, the membrane coverage determines the reduced volume $v$ and, thus, the membrane morphology.

\begin{figure*}[bth!]
\centering
\includegraphics[width=0.8\linewidth]{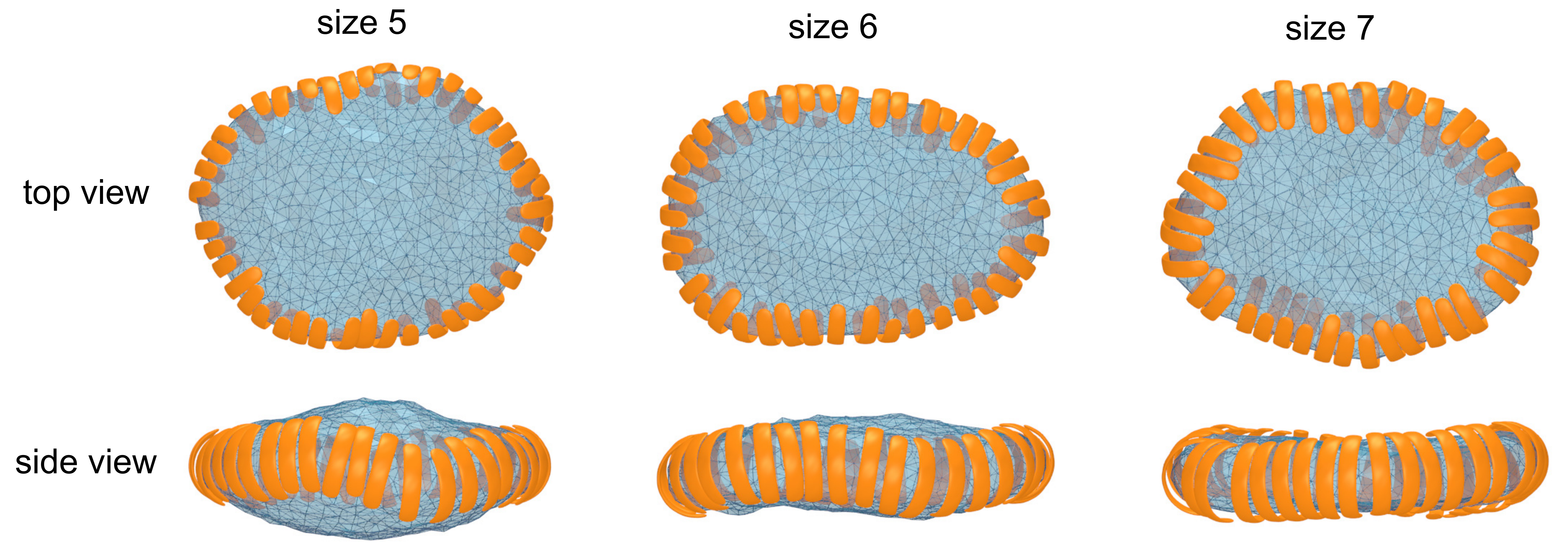}
\caption{ Exemplary disk-like morphologies for particles composed of 5 to 7 segments with an angle of $30\degree$ between adjacent segments and membrane coverages of $x = 0.28$, $0.31$, and $0.35$ (from left to right). The membranes consist of 2000 triangles.}
\label{figure-5}
\end{figure*}

All membrane morphologies of Figure  \ref{figure-2} are summarized in the diagram of Figure \ref{figure-3}. In this diagram, the mean induced angle is displayed versus the membrane coverage for the data points of Figure \ref{figure-2}. The mean induced angle varies from $52.5\degree$ for particles of size 3 with binding cutoff $\theta_c = 10\degree$ to  $169.7\degree$ for particles of size 7. Figure \ref{figure-3} illustrates that the threshold values of the membrane coverage above which membrane tubes are formed in our simulations is rather independent of the particle type. These threshold values range from about 40\% for particles of size 4 to 50\% for particles of size 3 and angle cutoff $\theta_c=10\degree$. Particles of size 5 to 7 induce membrane tubules at coverages above about 45\%, and disks at intermediate coverages below about 35\%. For particles of size 5, disk-like and tubular morphologies are separated by a somewhat larger gap in membrane coverage $x$, compared to particles of size 6 and 7. This gap arises because particles of size 5 are arranged in three lines along the tubes (see Figures \ref{figure-4} and S2), while elongated disks exhibit only two lines of particles at opposing sides (see Figures \ref{figure-5} and S3). Tubes with particles of size 5 thus cannot be generated by simple elongation of disks, in contrast to particles of size 6 and 7, which are arranged in two lines of particles along the tubes. Particles of size 4 induce tubular morphologies with four lines of particles along the tube.  At intermediate coverages, the particles lead to irregular, faceted morphologies with strongly curved membrane ridges covered by lines of particles, and weakly curved, uncovered  membrane segments in between these ridges (see Figure \ref{figure-2}, top). Particles of size 3 tend to align side by side along the tubules, but do not form continuous lines along the whole tubule. The ordering of particles of size 3 along the tubes is thus shorter ranged compared to larger particles. 

\begin{figure*}[hbt!]
\centering
\includegraphics[width=0.8\linewidth]{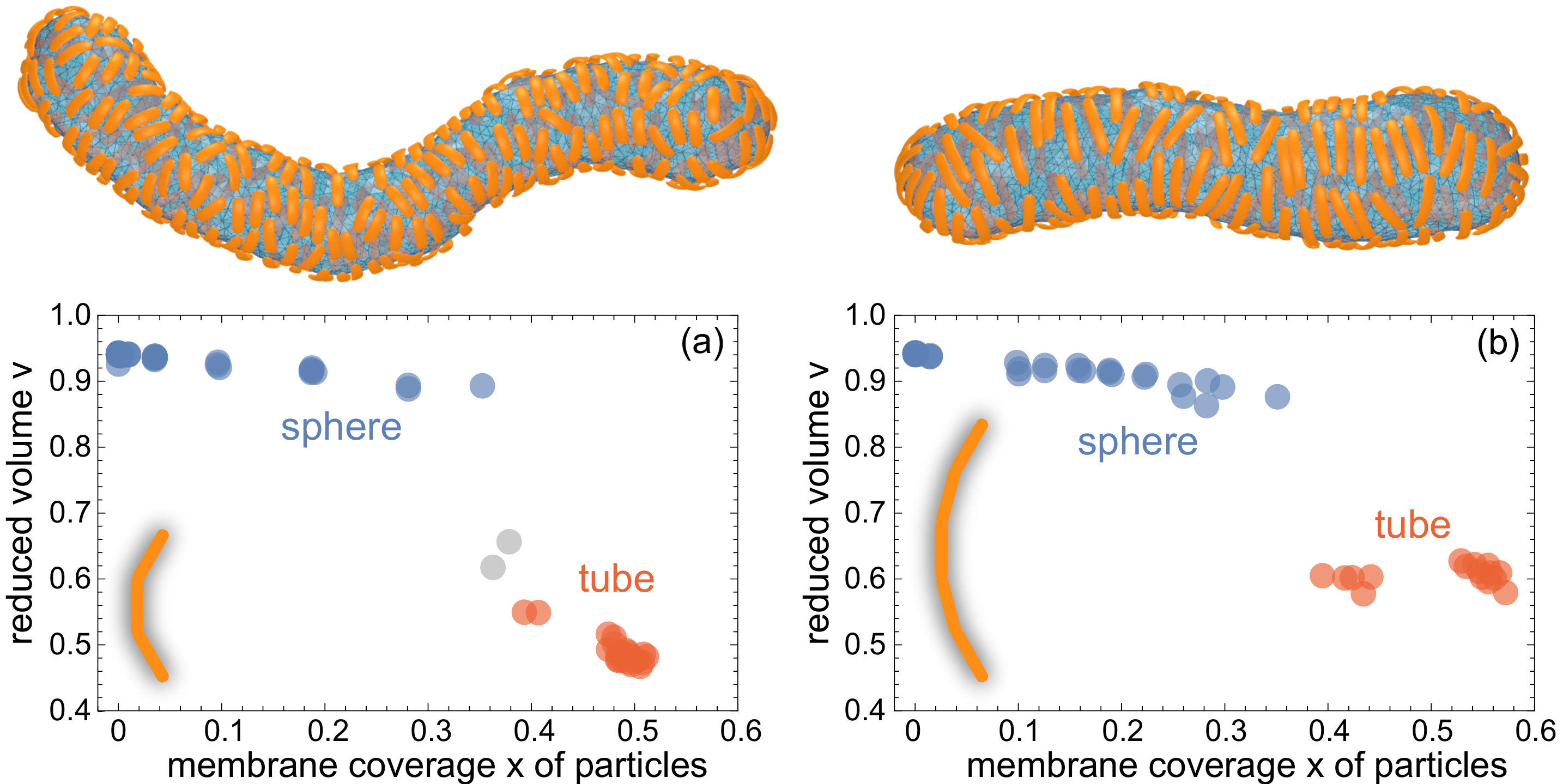}
\caption{(a) Reduced volume $v$ versus membrane coverage for membrane morphologies with 5120 membrane triangles and (a) particles composed of 3 segments with an angle of $30\degree$ between neighboring segments and (b) particles composed of 5 segments with an angle of $15\degree$ between neighboring segments. Spherical morphologies are indicated by blue points, tubular morphologies by red points. The two grey points in (a) correspond to intermediate morphologies. The data result from simulations with overall particle number $N=400$. To attain different coverages $x$, we have run simulations with adhesion energy $U = 3$, $3.5$, $4$, $4.5$, $5$, $5.5$, $6$, $6.5$, $6.6$, $6.8$, $7$, $7.2$, $7.4$, $7.5$, $7.6$, $7.8$, $8$, $8.2$, $8.4$, $8.5$, $8.6$, $9$, $10$, $11$, $12$, $13$, and $14\, k_B T$ in (a) and $U = 3$, $3.5$, $4$, $4.1$, $4.2$, $4.3$, $4.4$, $4.5$, $4.6$, $4.7$, $4.8$, $4.9$, $5$, $6$, $7$, $8$, $9$, $10$, $11$, $12$, and $13\, k_B T$ in (b). For each value of $U$, we have run 3 simulations starting from a spherical morphology and 2 to 3 simulations starting from a tubular morphology.  Only points from simulations with converged membrane coverage and reduced volume are included in the plots (see Methods). The mean induced angle of the particles in (b) is $56.1\degree$ and, thus, slightly larger than the mean induced angle of $52.5\degree$ of the particles in (a).
The particle coverage is $x = 0.48$ for the tubule shown to the left and $x = 0.44$ for the tubule to the right. 
}
\label{figure-6}
\end{figure*}

In Figure \ref{figure-6}, we compare simulation results for (a) particles composed of 3 segments with an angle of $30\degree$ between neighboring segments and (b) particles composed of 5 segments with an angle of $15\degree$ between neighboring segments. Both types of particles enclose the same arc angle of $60\degree$ between their terminal segments, but have different curvatures because of the different angles between their neighboring segments, and different sizes. 
The membrane in the simulations of Figure \ref{figure-6} is composed of 5120 triangles and is, thus, significantly larger than the membrane in the simulations of Figure \ref{figure-2}. For membranes composed of 2000 triangles as in Figure \ref{figure-2}, the more weakly curved particles of Figure \ref{figure-6}(b) do not induce a clear morphology transition from spherical to tubular because this smaller membrane size does not allow for sufficiently elongated spherocylinders that are clearly distinguishable from spheres (data not shown). For the larger membrane size of 5120 triangles, however, both types of particles of Figure \ref{figure-6} exhibit a rather sharp morphology transition from spherical to tubular at membrane coverages of about 0.37. This identical threshold value for the sphere-to-tubule transition illustrates that the overall membrane morphology is determined by the arc angle, which is identical for both types of particles, and not by the size or curvature of the particles. As expected, the more weakly curved particles of Figure \ref{figure-6}(b) induce thicker tubules for membrane coverages beyond the threshold value of about 0.37. For particles composed of 3 segments with an angle of $30\degree$ between neighboring segments, the threshold value for membranes with 5120 triangles is smaller than the threshold value of about 0.5 obtained for membranes with 2000 triangles (see Figure \ref{figure-2}(a)). The tubules of membranes with 2000 triangles induced by particles of 3 segments have a relatively small aspect ratio, i.e.\ a relatively small ratio of tube length and diameter, and dumbbell-like distortions for coverages $x$ close to the threshold value for tube formation (see Figure S3). The membrane size of 2000 triangles therefore is likely too small for reliable estimates of the sphere-to-tubule transition for these particles. The larger particles of Figure \ref{figure-2}, in contrast, induce tubules with significantly larger aspect ratios, compared to particles of size 3 (see Figures \ref{figure-4} and S3). The threshold values for tube formation obtained for these particles from Figures 2(c) to (f) therefore should be only weakly affected by the membrane size. 

\section*{Discussion and conclusions}

In this article, we have investigated the transitions between different membrane morphologies induced by arc-shaped particles { with purely repulsive direct particle-particle interactions}.  Our aim was to classify the membrane morphologies and to identify the particle properties that determine these morphologies. Arc-shaped particles can differ in their size, curvature, arc angle, adhesion energy, and overall number. Our central result is that the membrane morphologies induced by arc-shaped particles are determined by the arc angle and the membrane coverage of the particles.  In our model, the particles are described as segmented arcs that adhere to the triangulated membrane. The particle discretization thus is independent from the membrane discretization, and the membrane coverages obtained in our simulations are not affected or limited by the membrane discretization. In our simulations, the overall number of particles is larger than the number of adsorbed particles. Unbound particles thus constitute a particle reservoir in our simulations, which leads to rather sharp transitions and `pure' morphologies, as typical for simulations in a grand-canonical ensemble. We find that arc-shaped particles induce membrane tubules for membrane coverages larger than a threshold value of about 0.4, rather independent of their arc angle. At smaller coverages, particles with arc angles larger than about $120\degree$ induce disk-like membrane morphologies. These disk-like morphologies have characteristic membrane edges that connect the two opposing, nearly planar membrane segments of the disks. The disks are therefore `small-membrane equivalents' of the stacked, connected membrane sheets observed in the endoplasmic reticulum \cite{Terasaki13,Nixon16}. On the coarse-graining level of our model, the curvature generation of the particles is captured by induced curvature angles of the particles. Proteins may induce comparable curvatures and curvature angles by membrane adhesion, membrane insertion, or a combination of adhesion and insertion  \cite{McMahon15,Kozlov14,Baumgart11,Zimmerberg06}.

{ In our simulations, particles with arc angles of $60\degree$ induce a characteristic sphere-to-tubule transition. The membrane adopts a spherical shape for particle coverages below a transition value, and a tubular shape for coverages above the transition value. At coverages below the transition value, the particle arrangement 
on the spherical membranes is rather homogeneous, with only short-range order. Particles with arc angles of $90\degree$ and larger, in contrast, do not exhibit homogeneous arrangements on spherical membranes. Instead, these particles form linear aggregates at small membranes coverages in which the particles are aligned side-by-side. This alignment in lines are driven by indirect, membrane-mediated interactions between bound particles \cite{Weikl18,Phillips09,Reynwar07} because the direct particle-particle interactions are purely repulsive in our model. For particles with arc angles of $90\degree$, the linear aggregates at small membrane coverages lead to faceted morphologies, and to a rather continuous change from faceted morphologies to tubular morphologies with both four lines of particles for increasing membrane coverages (see also Figure S6). Particles with arc angles of $120\degree$, $150\degree$, and $180\degree$ form single closed linear aggregates at small membrane coverages in our simulations with excess particles. These closed linear aggregates lead to disk-like membrane morphologies. For particles with arc angles of $150\degree$ and $180\degree$, the disk-like morphologies change rather continuously into tubular morphologies with both two lines of particles along opposite membrane sites.  For particles with arc angles of $120\degree$, the transition from disk-like to tubular morphologies is discontinuous because the particles are arranged in three lines along tubes, which cannot form continuously from the single closed particle line of the disk-like morphologies. Overall, the transition from faceted or disk-like morphologies into tubular morphologies depend on the packing density of particles in the linear aggregates, and on the distances between the linear particle aggregates, which depends on the membrane coverage. It is plausible to assume that both the particle density along linear aggregates and the distances between linear aggregates for given membrane coverages do not depend on the overall membrane size in our simulations. The threshold coverage values for tube formation obtained from our simulations therefore should be independent of the membrane size, at least for particles with arc angles of $90\degree$ and larger. The sphere-to-tubule transition of our particles with arc angles of $60\degree$ in principle may depend on the membrane size, mainly because the curvature of the spherical membranes changes with increasing size. However, the rather large spherical membranes in the simulations of Figure \ref{figure-6} have a curvature that is already significantly smaller than the particle curvature. We therefore expect that the threshold values of the sphere-to-tubule transition obtained from Figure \ref{figure-6} do not change substantially with increasing membrane size. The tubules in our simulations have two spherical caps with a bending energy that corresponds to the bending energy of a sphere. The excess bending in the sphere-to-tubule transition therefore occurs in the cylindrical section of our tubules. 
}

{ Arc angles of $60\degree$ roughly correspond to the angle enclosed by BAR domain proteins such as the Arfaptin BAR domain and the endophilin and amphiphysin N-BAR domains \cite{Qualmann11,Masuda10}. Electron tomography images of membrane tubules induced by Bin1 N-BAR domain proteins show a rather loose protein arrangement with only short-ranged order (see Figure 5I of reference \cite{Daum16}), which is rather similar to the tubular morphologies obtained for our arc-shaped particles with arc angles of $60\degree$  (see Figures 4, S4, and 6). This similarity suggests that the observed rather loose arrangements of Bin1 N-BAR domains are likely dominated by membrane-mediated interactions. However, in contrast to our arc-shaped particles, the arrangements of BAR domain proteins in general can be affected by  specific protein-protein interactions. The highly ordered coats of CIP4 F-BAR domains \cite{Frost08}, endophilin N-BAR domains \cite{Mim12a}, and amphiphysin/BIN1 N-BAR domains \cite{Adam15} observed in electron microscopy appear to result from such specific protein-protein interactions. 3D reconstructions show that neighboring CIP4 F-BAR domains and amphiphysin/BIN1 N-BAR domains are arranged in parallel in these coats, but rather tip-to-side than side-to-side. These parallel tip-to-side arrangements in two-dimensional helical aggregates is clearly different from the linear, membrane-mediated aggregates observed in our simulations.  Fluorescence experiments of the membrane tubulation of giant unilamellar vesicles (GUVs) by endophilin N-BAR domains indicate an onset of tubulation at endophilin coverages of about 10\% on the vesicles for low membrane tensions \cite{Shi15}. The endophilin coverage along the tubules has not been determined in these experiments. However, the tubulation of endophilin N-BAR domain observed in these fluorescence experiments is clearly different from the sphere-to-tube transition induced by our particles with arc angles of $60\degree$. In this transition, membrane spheres with particle coverages close to $40\degree$ change into tubes with about the same coverage.  In fluorescence experiments with pre-formed tubules pulled with optical tweezers from GUVs, the membrane coverages $44\pm 27\%$ of endophilin N-BAR domains and $37\pm 9\%$ of $\beta$2 centaurin BAR domains have been measured along the tubules \cite{Simunovic16}. }
Sparse but regular arrangements on membrane tubules have been recently observed for dynamin-amphiphysin complexes \cite{Takeda18}. 

The membrane-mediated side-by-side alignment of particles in our simulations is also prominent in the time sequences of particle-induced morphology changes from spherical to tubular  (see Figures S4, S5, and S5).  Attractive membrane-mediated pair interactions of arc-shaped particles that are oriented side-by-side have been previously found by energy minimization \cite{Schweitzer15a}. In molecular dynamics (MD) simulations with a coarse-grained molecular model of N-BAR domains proteins on DLPC lipid vesicles, in contrast, a tip-to-tip alignment of proteins has been observed \cite{Simunovic13,Simunovic17}. 
 
Morphologies with tubular and disk-like membrane segments have been previously observed in different elastic-membrane models \cite{Ayton07,Ayton09,Ramakrishnan13,Tourdot14,Noguchi15,Noguchi16,Noguchi16b}. In recent models, curvature-inducing particles have been described as nematic objects embedded on the vertices of a triangulated membrane \cite{Ramakrishnan13,Tourdot14}, or as curved chains of beads embedded in a two-dimensional sheet of beads that represents the membrane  \cite{Noguchi15,Noguchi16}.  These models have been investigated for a constant number of membrane-embedded particles. For such a canonical ensemble of embedded particles, membrane-mediated interactions between the particles lead to particle aggregation and to membrane morphologies with particle-dense, strongly curved membrane segments and particle-free, weakly curved segments. The membrane coverages in the particle-dense regions are affected by the membrane discretization in these models. Curved chains of beads embedded in a sheet of beads, for example, tend to contact each other side by side  \cite{Noguchi15,Noguchi16}. The membrane coverage in these dense particle regions then depends on the size of the discrete membrane beads. In agreement with our simulation results, the curved chains of beads form several lines along tubules, depending on their chain length \cite{Noguchi15}.

In our disk-like morphologies, the arc-shaped particles stabilize a highly curved, closed edge that connects two stacked, weakly curved membrane segments. The thickness of the membrane disks decreases with the particle size and, thus, with the induced angle of the particles (see Figures S7, \ref{figure-5}, and S3), in agreement with energy minimization for ideal disks \cite{Schweitzer15b}. The thickness and  coverage of the disks in our model are limited by the membrane area. For larger membrane area A, disk-like morphologies exhibit smaller overall coverages because the curved, particle-covered edge of the disks increases only proportional to $\sqrt{A}$. In general, stacked membranes with lateral extensions that are significantly larger than their separation repel each other sterically because of membrane shape fluctuations \cite{Helfrich78,Lipowsky86}. Therefore, stacked membranes of large area are presumably stabilized by additional attractive interactions between the membranes.

The membrane morphologies in our model result from an intricate interplay of the bending free energy of the membrane and the overall adhesion free energy of the particles. The membrane in our model is tensionless. In general, the bending energy dominates over a membrane tension $\sigma$ on length scales smaller than the characteristic length $\sqrt{\kappa/\sigma}$, which adopts values between 100 and 400 nm for typical tensions $\sigma$ of a few $\mu\text{N}/\text{m}$ \cite{Simson98,Popescu06,Betz09} and typical bending rigidities $\kappa$ between $10$ and $40$ $k_B T$ where $k_B T$ is the thermal energy \cite{Nagle13,Dimova14}. Our results thus hold on length scales smaller than this characteristic length. In contrast, the overall membrane morphology on length scales larger than $\sqrt{\kappa/\sigma}$ depends on the membrane tension \cite{Lipowsky13,Shi15,Simunovic15}.

\section*{Author Contributions}

FB and TW designed the research. FB carried out all simulations. FB and TW analyzed the data and wrote the article. 

\section*{Acknowledgments}

Financial support from the Deutsche Forschungsgemeinschaft (DFG) {\em via} the International Research Training Group 1524 ``Self-Assembled Soft Matter Nano-Structures at Interfaces" is gratefully acknowledged.



\clearpage
\beginsupplement

\begin{figure*}[tbh]
{\bf\large SUPPLEMENTARY FIGURES}\\[3cm]
\begin{center}
\resizebox{1.1\columnwidth}{!}{\includegraphics{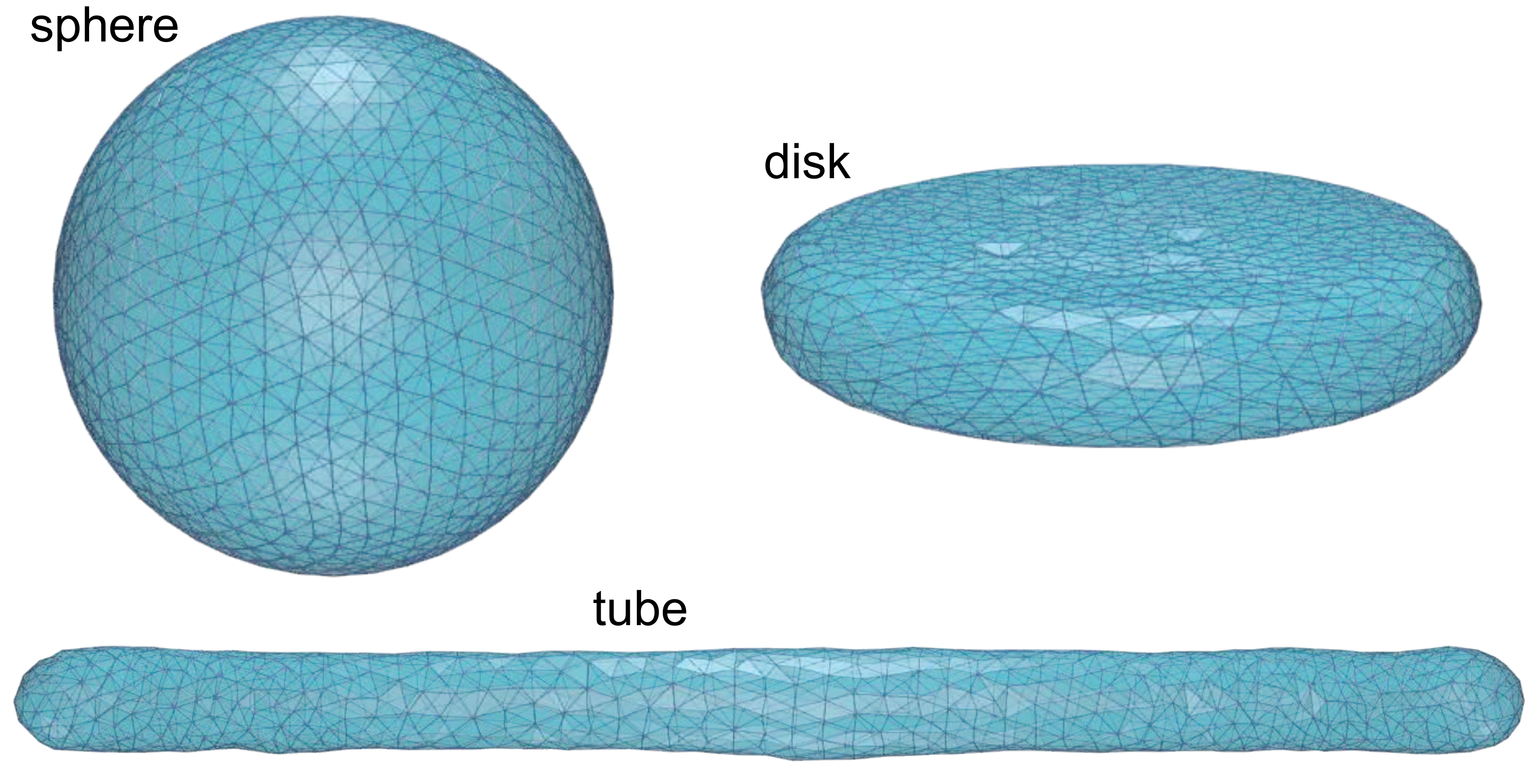}}
\end{center}
\caption{Initial spherical, disk-like, and tubular membrane shapes of our MC simulations.}
~\\[8cm]
\label{figure-S-initial}
\end{figure*}

\begin{figure*}[h]
\begin{center}
\resizebox{1.2\columnwidth}{!}{\includegraphics{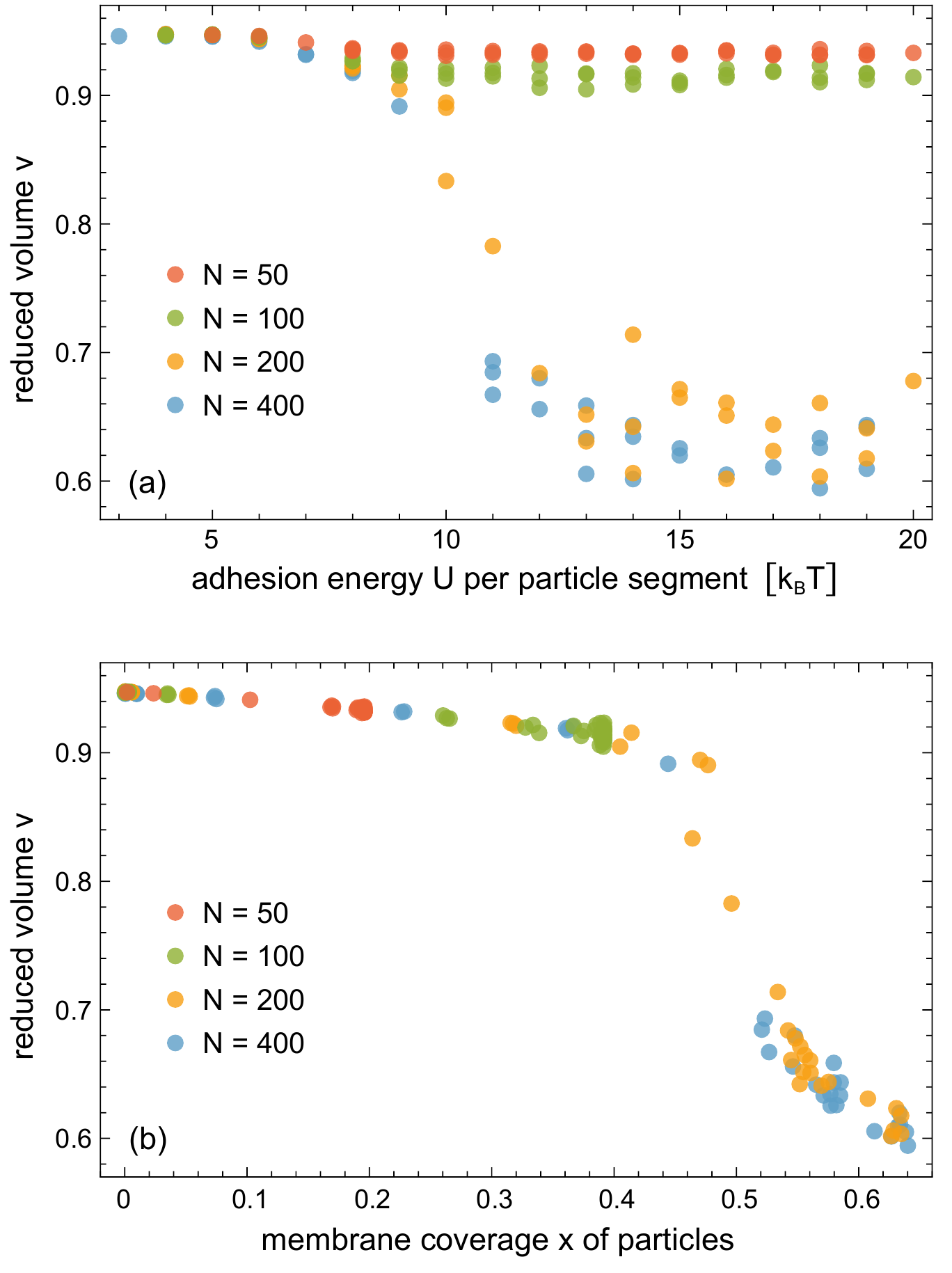}}
\end{center}
\caption{Reduced volume $v$ of the membrane vesicles versus (a) adhesion energy $U$ per particle segment and (b) membrane coverage $x$ for particles composed of 3 segments at different overall particle numbers $N$. The data for the overall particle numbers $N = 200$ and $400$ are identical to the data shown in Figure 2(a). For the particle numbers $N = 50$ and $100$, the membrane coverage $x$ is not sufficiently high to induce membrane tubules with $v\lesssim 0.7$ even for large adhesion energies $U$. As a function of $x$, all data collapse onto a single curve, which illustrates that the reduced volume $v$ and, thus, the membrane morphology, is determined by the membrane coverage $x$ of the particles. This coverage of membrane-bound particles depends both on the overall particle number $N$ and on the adhesion energy $U$ per particle segment. 
} 
\label{figure-S-thickness}
\end{figure*}

\begin{figure*}[t]
\hspace*{-1cm}
\resizebox{2.2\columnwidth}{!}{\includegraphics{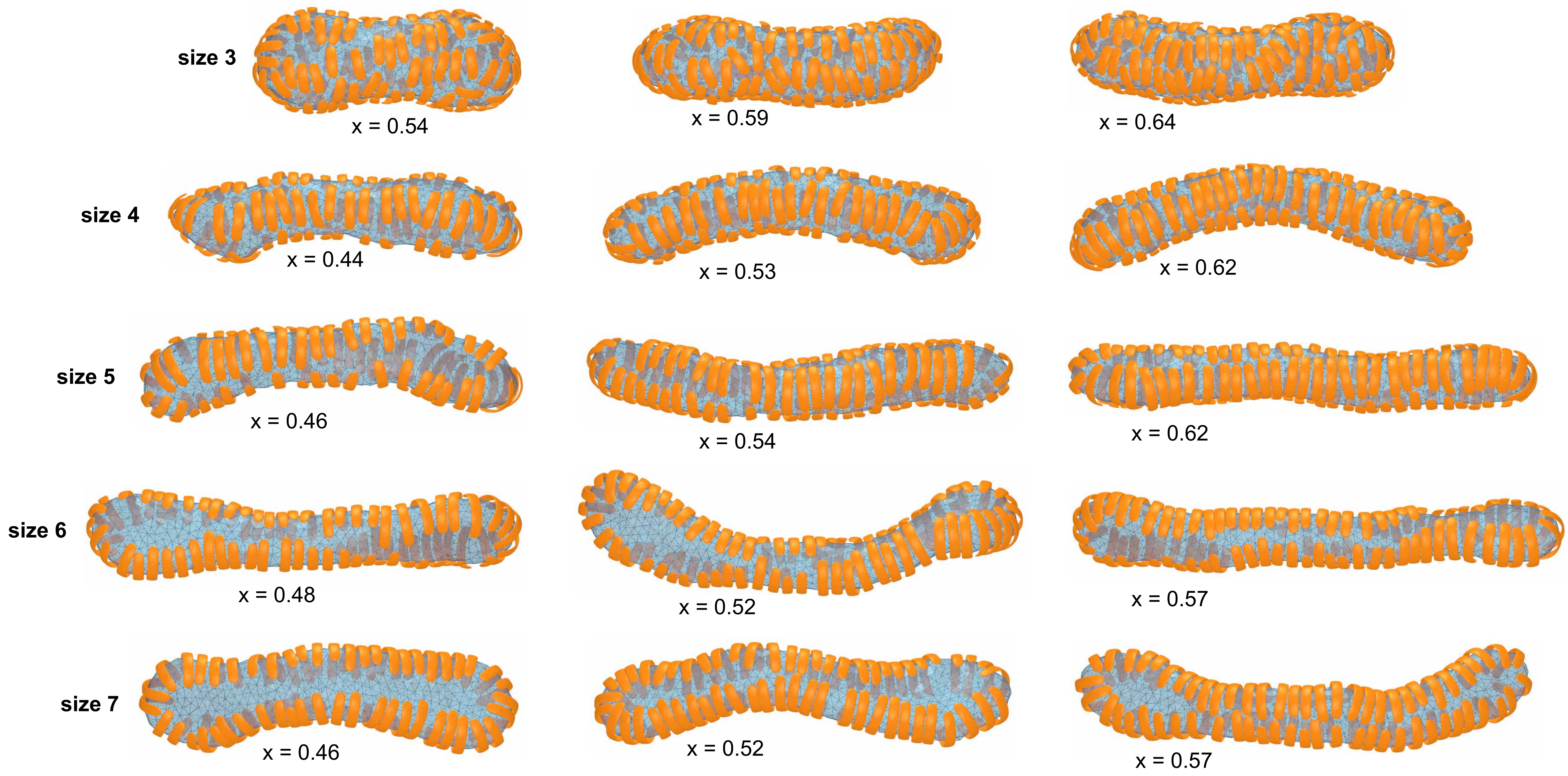}}
\caption{Exemplary tubular morphologies for particles of size 3 to 7 at different membrane coverages $x$.}
\label{figure-S-more-tubes}
\end{figure*}

\begin{figure*}[t]
\begin{center}
\resizebox{1.6\columnwidth}{!}{\includegraphics{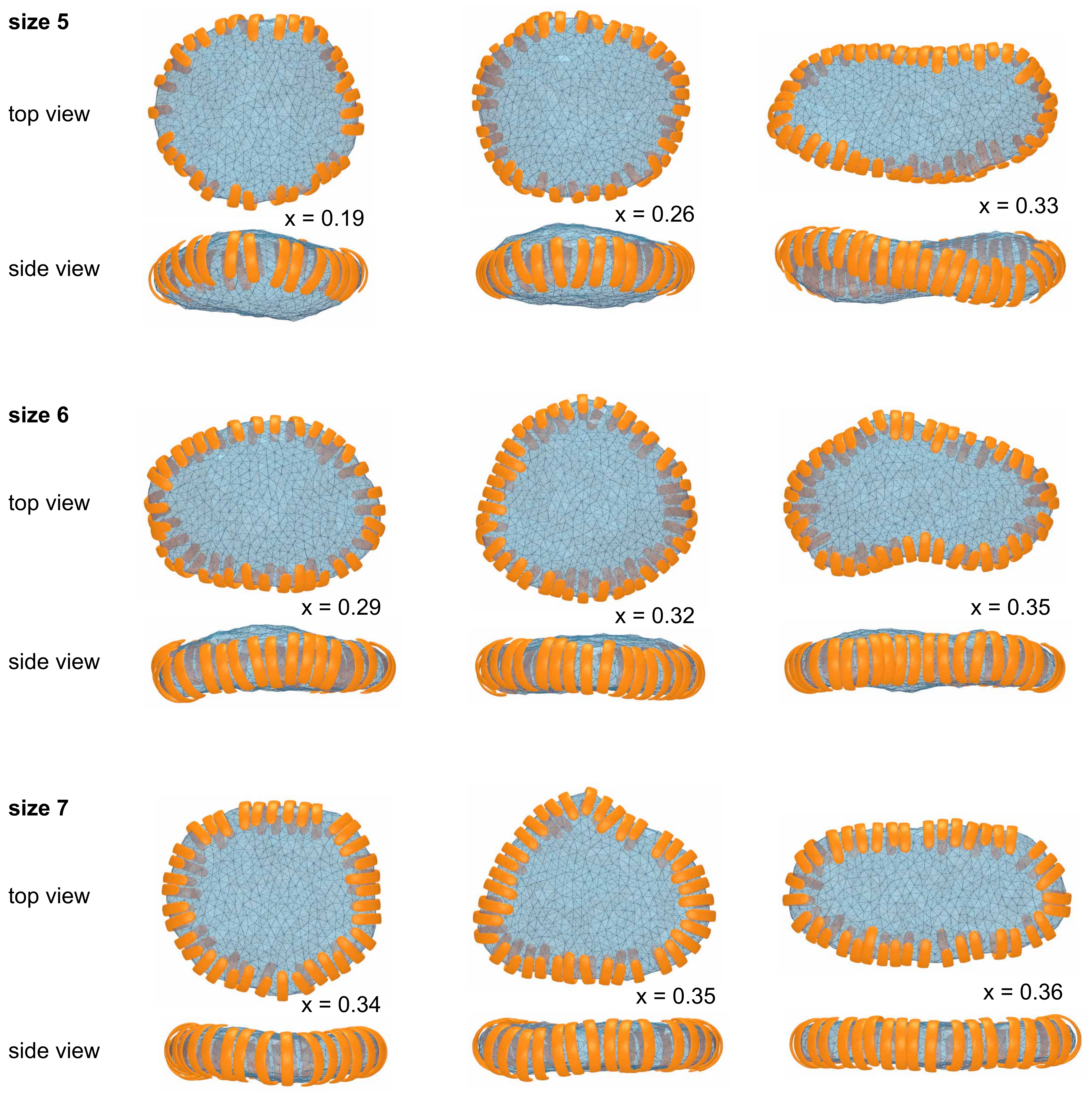}}
\caption{Exemplary disk-like morphologies for particles of size 5 to 7 at different membrane coverages $x$. }
\end{center}
\label{figure-S-more-disks}
\end{figure*}

\begin{figure*}[t]
\resizebox{2\columnwidth}{!}{\includegraphics{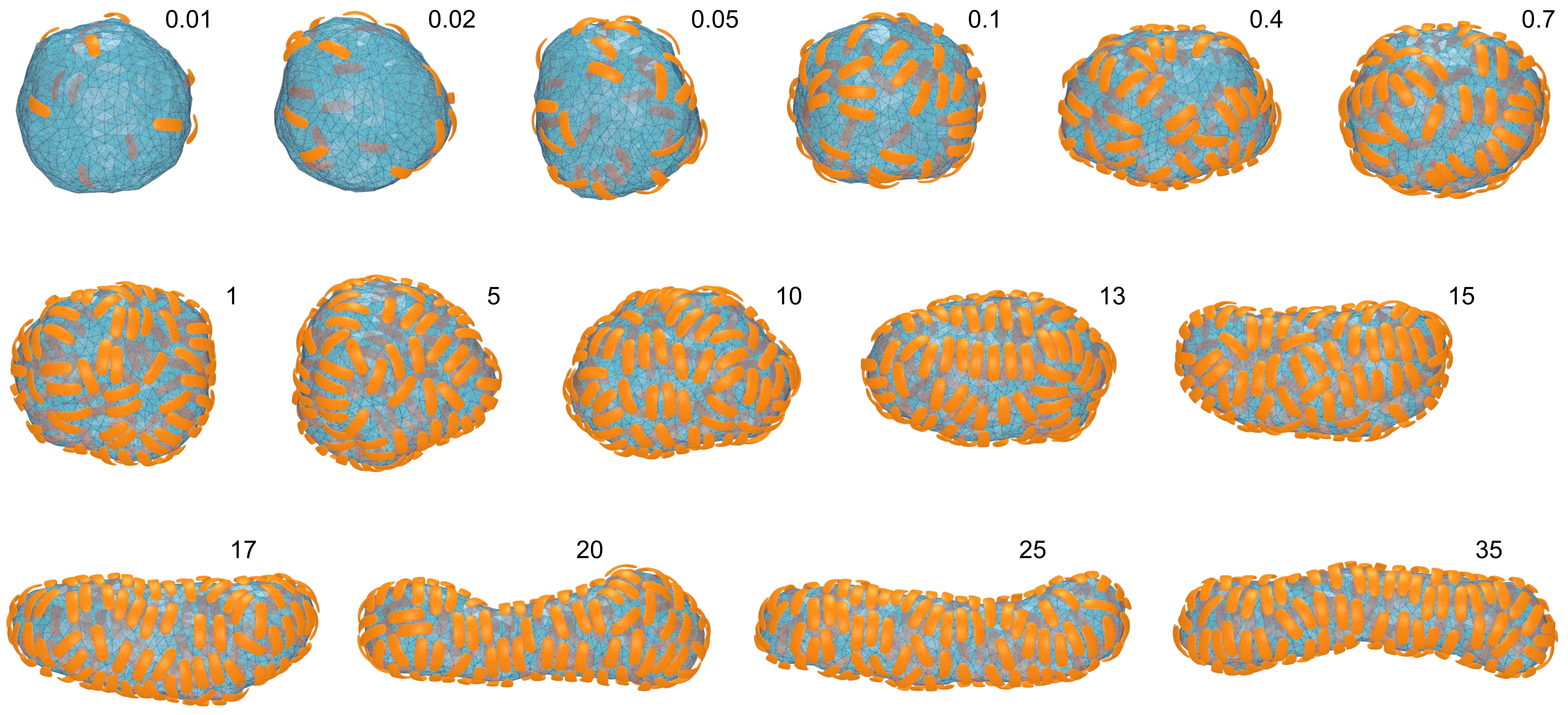}}
\caption{Time sequence of a morphology change from spherical to tubular induced by arc-shaped particles with 3 segments. The numbers indicate simulation times in units of $10^6$ MC steps per membrane vertex. At time 0, the membrane has the initial spherical shape depicted in Figure \ref{figure-S-initial}, and all particles are unbound.  In this simulation, the adhesion energy per particle segment is  $U = 18 k_B T$, the cutoff angle for binding is $\theta_c = 10\degree$, and the total number of particles is 400. Only bound particles are shown in the MC snapshots. 
}
\label{figure-S-t3}
\end{figure*}

\begin{figure*}[t]
\resizebox{2\columnwidth}{!}{\includegraphics{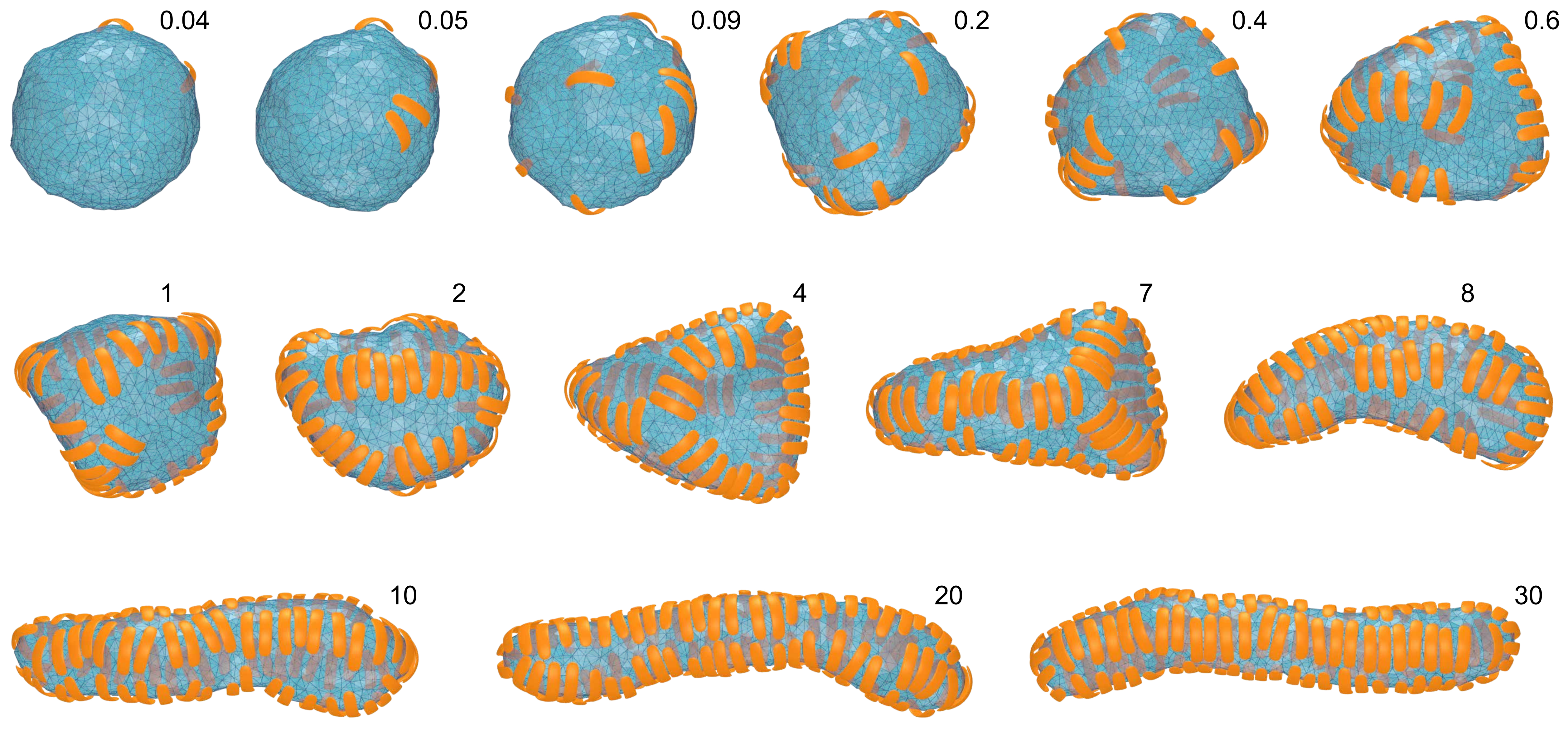}}
\caption{Time sequence of a morphology change from spherical to tubular induced by arc-shaped particles with 4 segments. The numbers indicate simulation times in units of $10^6$ MC steps per membrane vertex. At time 0, the membrane has the initial spherical shape depicted in Figure \ref{figure-S-initial}, and all particles are unbound.  In this simulation, the adhesion energy per particle segment is  $U = 18 k_B T$, and the total number of particles is 400. 
 }
\label{figure-S-t4}
\end{figure*}

\begin{figure*}[t]
\resizebox{2\columnwidth}{!}{\includegraphics{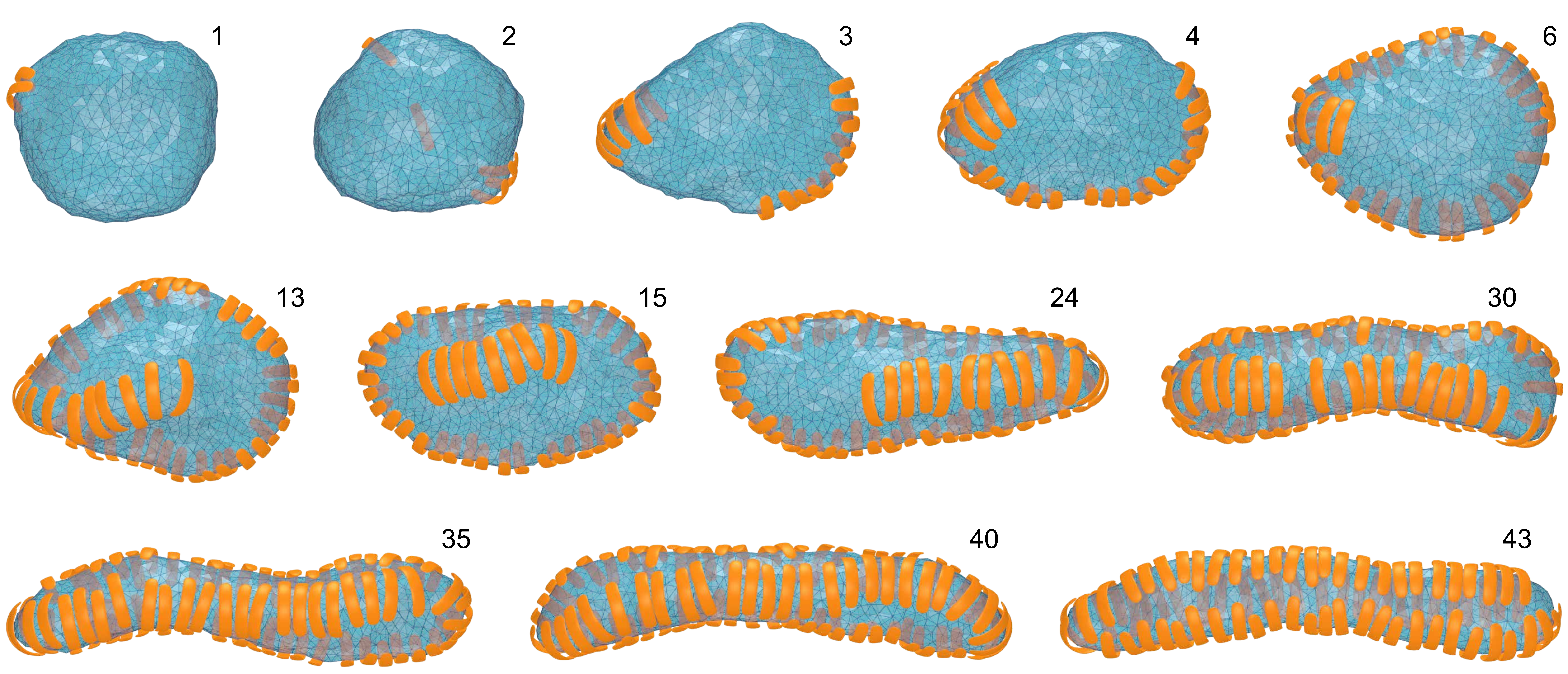}}
\caption{ Time sequence of a morphology change from spherical to tubular induced by arc-shaped particles with 5 segments. The numbers indicate simulation times in units of $10^6$ MC steps per membrane vertex. At time 0, the membrane has the initial spherical shape depicted in Figure \ref{figure-S-initial}, and all particles are unbound.  In this simulation, the adhesion energy per particle segment is  $U = 12 k_B T$, and the total number of particles is 400. }
\label{figure-S-t5}
\end{figure*}

\begin{figure*}[t]
\begin{center}
\resizebox{1.2\columnwidth}{!}{\includegraphics{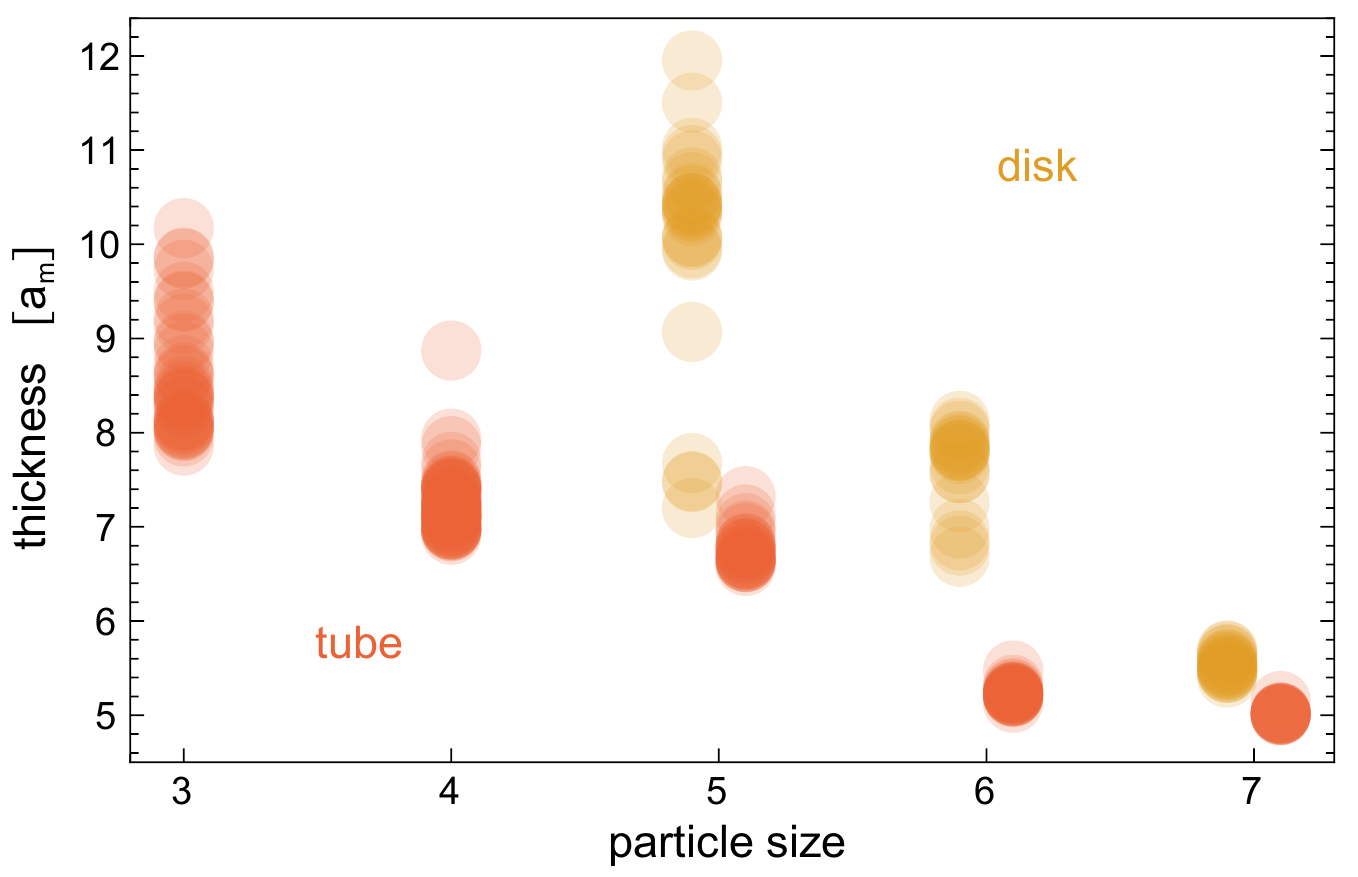}}
\end{center}
\caption{Thickness of disks and tubes in units of the minimum edge length $a_m$ of the triangulated membrane. To determine the disk thickness, each membrane vertex is connected to an opposing vertex such that the line between the vertices is closer to the center of mass of the membrane than lines connecting the vertix to other vertices. The disk thickness is defined as the minimum length of all lines between opposing pairs of vertices. To determine the tube thickness, we first project all vertices on the axis of inertia parallel to the tube and select those vertices for which the distance to the center of mass along this axis  is smaller than 2.3 $a_m$.  These vertices form a central segment of the tube. We then calculate the tube thickness as the thickness of this central segment in analogy to the disk thickness. Each data point in this figure corresponds to a data point in Figure 2.} 
\label{figure-S-thickness}
\end{figure*}

\end{document}